\documentclass[11pt,a4paper]{article}

\usepackage[colorlinks=true, linkcolor=black!50!blue, urlcolor=blue, citecolor=blue, anchorcolor=blue]{hyperref}
\usepackage[font=small,labelfont=bf,margin=0mm,labelsep=period,tableposition=top]{caption}
\usepackage[a4paper,top=3cm,bottom=2.5cm,left=2.5cm,right=2.5cm,bindingoffset=0mm]{geometry}

\usepackage{graphicx}
\usepackage{float}
\usepackage{afterpage}
\usepackage{epsfig,cite}
\usepackage{amssymb}
\usepackage{amsmath}
\usepackage{multirow}
\usepackage{url}
\usepackage{xcolor}
\usepackage{float}
\usepackage{afterpage}

\usepackage{url}

\usepackage{booktabs}
\usepackage{tikz}
\usetikzlibrary{arrows,backgrounds,snakes}
\usetikzlibrary{decorations.markings}
\usetikzlibrary{decorations.pathmorphing,decorations.markings}
\usepackage[compat=1.1.0]{tikz-feynhand}
\usetikzlibrary{shadows,arrows,intersections}
\usepackage{enumitem}
\usepackage{hyperref}
\usepackage{cite}
\usepackage{simpler-wick}

\newcommand{\ie}{{\em i.e.}}
\newcommand{\msbar}{$\overline{\mathrm{MS}}$}

\numberwithin{equation}{section}
\numberwithin{figure}{section}
\numberwithin{table}{section}

\usepackage{tabularx}
\newcolumntype{C}[1]{>{\centering\arraybackslash}p{#1}}

\begin{document}
\newgeometry{top=1.5cm,bottom=1.5cm,left=2.5cm,right=2.5cm,bindingoffset=0mm}

\vspace*{.2cm}

\begin{center}
{\Large \bf Notes on lattice observables for parton distributions:\\[0.2cm]
{\Large nongauge theories}}
\vspace{1.4cm}

 {\small
Luigi Del Debbio$^{1}$ 
Tommaso Giani$^{1}$ and
Christopher J. Monahan$^{2,3}$
}

\vspace{1.0cm}
 
{\it \small 
         ~$^1$ The Higgs Centre for Theoretical Physics, The University of Edinburgh,\\
  Peter Guthrie Tait Road, Edinburgh EH9 3FD, United Kingdom\\[0.1cm]
  ~$^2$ Physics Department, College of William and Mary, Williamsburg, VA 23187, United States\\
  ~$3$ Thomas Jefferson National Accelerator Facility, Newport News, Virginia 23606, USA
}

\vspace{1.0cm}
{\bf \large Abstract}
\end{center}

{\noindent We review recent theoretical developments concerning the definition and the renormalization of equal-time correlators that can be computed on the lattice and related to Parton Distribution Functions (PDFs) through a factorization formula. We show how these objects can be studied and analyzed within the framework of a nongauge theory, gaining insight through a one-loop computation. We use scalar field theory as a playground to revise, analyze and present the main features of these ideas, to explore their potential, and to understand their limitations for extracting PDFs. We then propose a framework that would allow to include the available lattice QCD data in a global analysis to extract PDFs.
}

\clearpage

\section{Introduction}
\label{sec:introduction}

In recent years, there has been a significant effort within the lattice
community to compute specific equal-time correlators that can be directly
related to Parton Distribution Functions (PDFs). PDFs describe the longitudinal
structure of nucleons in terms of their partonic constituents. They are
inherently non-perturbative quantities, which can be extracted from data using
so-called factorization theorems. Given the central role of PDFs in the analysis
of experimental data at hadronic colliders, it would be highly beneficial to be
able to use lattice QCD to determine these crucial ingredients in our current
understanding of nucleon structure. Quasi-PDFs\footnote{Quasi-PDFs are one example of the more general LaMET formalism \cite{Ji:2014gla,Ji:2020ect}, but here we focus on the collinear $x$-dependent distributions.} and pseudo-PDFs were introduced in
Refs.~\cite{PhysRevLett.110.262002, Radyushkin:2017cyf}, and since then numerous
publications have appeared, addressing the main theoretical issues for these
approaches. For recent reviews, we refer the reader to
Refs.~\cite{DelDebbio:2018siw,Monahan:2018euv,Zhao:2018fyu,Cichy:2018mum,
Radyushkin:2019mye,Ji:2020ect,Lin:2020rut}. This program has often been referred to as the ``first
principles computation of PDFs'', generating different reactions among the
lattice and high-energy physics communities: on the one hand it has been
welcomed with enthusiasm, triggering several dedicated studies; on the other
hand it has been criticized in Refs.~\cite{Rossi:2017muf,Rossi:2018zkn} on the
basis that equal-time correlators do not give access to the full
non-perturbative PDF. Both reactions are healthy and show the importance of the
original proposal in~\cite{PhysRevLett.110.262002}. This criticism mentioned
above has, in turn, been addressed in
Refs.~\cite{Radyushkin:2018nbf,Karpie:2018zaz}. Given the increasing number of
lattice calculations, there is a need to revise and clarify the main conceptual
questions: that is, how do we extract information on PDFs from quasi- and
pseudo-PDFs, and what is the interplay between quasi- and pseudo-PDFs with
experimental data?

In this paper we study these topics in the context of a renormalizable scalar
theory. Scalar field theory is a valuable model for understanding the essential
theoretical issues in a simple framework, as shown in the pioneering study of
PDFs by Collins in Ref.~\cite{Collins:1980ui}. We follow the ideas presented
there, which we extend to account for quasi- and pseudo-PDFs. Our aim is to
investigate, clarify and highlight some subtle points using scalar field theory
as a simple playground, and to assess how the lattice QCD results that are
currently available can be used to extract PDFs. 

We will consider a massive scalar field theory, in $d=6$ dimensions, with a
$\phi^3$ interaction term, whose bare Lagrangian $\mathcal{L}$ is given by
\begin{align}
    \label{eq:Lagrangian}
    \mathcal{L} = \frac{1}{2}\left(\partial\phi\right)^2 
    - \frac{m^2}{2} \phi^2 - \frac{g}{3!} \phi^3.
\end{align}
Working within this model allows us to analyze the conceptual framework for
quasi- and pseudo-PDFs in a clean and straightforward way, avoiding
complications associated with QCD that are unnecessary for understanding the
basics of these approaches. We focus in particular on the matrix element of a
field bilinear between ``nucleon'' states: 
\begin{align}
    \label{eq::ME}
    \mathcal{M} = 
    \langle P | \phi\left(z\right) \phi\left(0\right) | P \rangle\, ,
\end{align}
when the separation $z$ between the fields is either light-cone like, $z^2=0$,
or purely spatial, $z^2=-z_3^2$. In the first case, we obtain the matrix element
that underlies the formal definition of collinear PDFs~\cite{Collins:1980ui,
Collins:1981uw}, which are obtained as the Fourier transform along a light-cone
direction of the matrix element in Eq.\eqref{eq::ME}~\footnote{The field
bilinear needs to undergo a proper renormalization, which we explore in detail in
this paper. }:
\begin{align}
    \label{eq:barePDF}
    f(x) = xP^+ \int \frac{dz^-}{2\pi}\, e^{-i xP^+ z^-}\,
    \langle P | \phi\left(z\right) \phi\left(0\right) | P \rangle\, ,
\end{align}
where $P^+$ and $z^-$ are the usual light-cone coordinates of the four-vectors
$P$ and $z$ respectively. In the second case we obtain an equal-time correlator
that can be computed on the lattice. We address the problem of the
renormalization of these quantities and study the relation between them at one
loop in perturbation theory, both in position and momentum space. As we shall
see, the main features of the computation are the same as in QCD. This allows us
to understand easily the main concepts, relations and limitations of the quasi-
and pseudo-PDF approaches. With a clear picture of the theoretical background
and of what is currently available in the literature, we then propose a general
framework to extract collinear PDFs from the available lattice data, based on
the optimization of a parametric form of the PDFs. The implementation of such
approach has been started in Ref.~\cite{Cichy:2019ebf} within the {\tt NNPDF}
environment, using the same strategy that is commonly used to extract PDFs from
data for experimental observables.

We address, in turn, a number of questions that have been raised in the context of
QCD, and analyze the lessons that we can draw from the scalar model. 

First we discuss issues that are related to the analysis of ultraviolet (UV)
divergences of the bilinear operator and their subtraction through the
renormalization process. In particular in Sec.~\ref{sec:light-cone} we perform
the computation of $\mathcal{M}$ in the case of a light-cone separation,
recovering the results of Ref.~\cite{Collins:1980ui} through a position space
calculation. In Sec.~\ref{sec:spatial-separtaion} we perform the same exercise
outside the light-cone, choosing a purely spatial separation between fields, and
we discuss the main differences with respect to the light-cone case. 

In both cases, we define quantities that are free of divergences when the
regulator is removed, and then focus on the relation between light-cone and
equal-time correlators. In Sec.~\ref{sec:factorization} we work out this
relation explicitly at one loop in perturbation theory, and analyze the limits
leading to a factorization theorem, in both position and momentum space, and
in Sec.~\ref{sec:flow} we extend the discussion to include smeared equal-time correlators.  

In Sec.~\ref{sec:conclusion} we summarize, discuss how these ideas can be used
in a fitting framework to extract PDFs, and draw our conclusions. The work is
supplemented with a number of appendices containing the technical details of the
computations and addressing the objections raised in
Refs.~\cite{Rossi:2017muf,Rossi:2018zkn}.

\section{Light-cone separation}
\label{sec:light-cone}

As stressed in Ref.~\cite{Radyushkin:2017cyf}, the matrix element defined in
Eq.~\eqref{eq::ME} is a function of the Lorentz invariants $z^2$ and $\nu =
P\cdot z$, the ``Ioffe time'', so that we can write
$\mathcal{M}=\mathcal{M}\left(\nu,z^2\right)$. In this section we focus on the
perturbative renormalization of $\mathcal{M}\left(\nu,z^2\right)$ at the
one-loop level, in the light-cone separation case, $z^2=0$. 
We work in perturbation theory, denoting the bare field of our theory as $\phi$,
and we consider partonic matrix elements
\begin{align}
\label{eq::MEpartonic}
        \widehat{\mathcal{M}}\left(\nu,z^2\right) = \langle p | \phi\left(z\right)\phi\left(0\right)  | p \rangle
\end{align}
between on-shell quark states with four-momentum $p$, with $p^2 =
m_\mathrm{pole}^2$. Throughout this calculation, we denote partonic quantities
with a ``hat'', while the lower-case $p$ refers to the momentum of the parton.
In what follows the Lorentz invariant $\nu$ is defined as $\nu=p\cdot z$.
Restricting ourselves to the case for which $z_0 \geqslant 0 $, we have
\begin{align}
\label{eq::LSZ}
        \widehat{\mathcal{M}}&\left(\nu, z^2\right)=\langle p |T\left[\phi\left(z\right)\phi\left(0\right)\right]|p\rangle  \nonumber \\
        & = \lim_{p^2\rightarrow m^2_{\text{pole}}}\left(p^2-m^2_{\text{pole}} + i\epsilon\right)^2 
        \int dz_1\,dz_2 \,e^{-ip\cdot z_1}e^{ip\cdot z_2}\,
        \langle 0 | T\left[\phi\left(z\right)\phi\left(0\right)\phi\left(z_1\right)\phi\left(z_2\right)\right]|0\rangle\, ,
\end{align}
where $m_\mathrm{pole}^2$ is defined by the location of the pole in the scalar
propagator, and can be computed at each order in perturbation theory. At tree
level we have $m_\mathrm{pole}^2 = m^2$, while in general $m_\mathrm{pole}^2 - m^2 = \mathcal{O}\left(g^2\right)$.

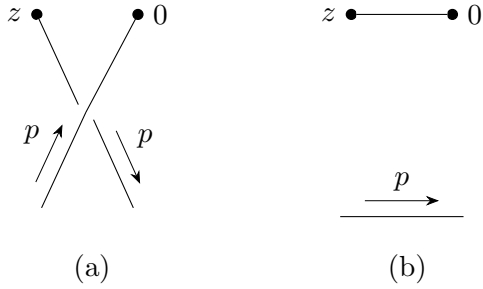
\begin{figure}[h]
        \centering
        \begin{tikzpicture}
                \renewcommand{\feynhandtopsepcolor}{white}
                \setlength{\feynhandtopsep}{8pt} 
              \begin{feynhand} 
                \vertex [dot] (x) {};
                \vertex [left=0.cm of x] {$z$};
                \vertex [dot] [right=1.2cm of x] (y) {};
                \vertex [right=0.cm of y] {$0$};
                \vertex [below=2.5cm of x] (p) {};
                \vertex [below=2.5cm of y] (pprime) {};
                \vertex [below right=3.0cm and 0.3cm of x] {(a)};
                \vertex [below right=1.25cm and 0.6cm of x] (v);
                \vertex [above left=.1cm and 0.1cm of v] (v1);
                \vertex [below right=.1cm and 0.1cm of v] (v2);
                \propag [top] (x) to (v1);
                \propag [top, mom={[arrow shorten=0.2]$p$}] (v2) to (pprime);
                \propag [mom={[arrow shorten=0.2]$p$}] (p) to (v);
                \propag [] (v) to (y);
              \end{feynhand}
            \end{tikzpicture}
          \hspace{1.5truecm}
          \begin{tikzpicture}
            \renewcommand{\feynhandtopsepcolor}{white}
            \setlength{\feynhandtopsep}{8pt} 
          \begin{feynhand} 
            \vertex [dot] (x) {};
            \vertex [left=0.cm of x] {$z$};
            \vertex [dot] [right=1.2cm of x] (y) {};
            \vertex [right=0.cm of y] {$0$};
            \vertex [below left=2.5cm and 0.1cm of x] (p) {};
            \vertex [below right=2.5cm and 0.1cm of y] (pprime) {};
            \vertex [below right=3.0cm and 0.3cm of x] {(b)};
            \vertex [below right=1.25cm and 0.6cm of x] (v);
            \vertex [above left=.1cm and 0.1cm of v] (v1);
            \vertex [below right=.1cm and 0.1cm of v] (v2);
            \propag [top] (x) to (y);
            \propag [top, mom={[arrow shorten=0.2]$p$}] (p) to (pprime);
          \end{feynhand}
          \end{tikzpicture} 
	\vspace*{5mm}
	\caption{Contractions that are not considered in the present discussion. Diagram (a) is excluded when considering contributions proportional to $\exp(-ip\cdot z)$, while diagram (b) cancels when looking at the connected correlator.}
	\label{fig:1}
\end{figure}

When computing the 4-point function entering Eq.~\eqref{eq::LSZ}, we will not
consider diagrams like those in Fig.~\ref{fig:1}. Following
Ref.~\cite{Collins:1980ui}, we are only interested in the contribution
proportional to $\exp(-ip\cdot z)$, and therefore discard topologies like the
one in diagram (a). Diagram (b) is removed by considering the connected
contribution only. 

\setlength{\feynhanddotsize}{1.3truemm}

Therefore the only Feynman diagrams contributing to Eq.~\eqref{eq::LSZ} up to one-loop order are those shown in Fig.~\ref{fig:2}.
\begin{figure}[h]
        \centering
    \begin{tikzpicture}
    \begin{feynhand} 
      \vertex [dot] (x) {};
      \vertex [left=0.cm of x] {$z$};
      \vertex [dot] [right=1.2cm of x] (y) {};
      \vertex [right=0.cm of y] {$0$};
      \vertex [below=2.5cm of x] (p) {};
      \vertex [below=2.5cm of y] (pprime) {};
      \vertex [below right=3.0cm and 0.3cm of x] {(a)};
      \propag [revmom={[arrow shorten=0.3]$p$}] (x) to (p);
      \propag [revmom'={[arrow shorten=0.3]$p$}] (pprime) to (y);
    \end{feynhand}
  \end{tikzpicture}
\hspace{.5truecm}
   \begin{tikzpicture} 
     \begin{feynhand} 
     \vertex [dot] (x) {};
     \vertex [left=0.cm of x] {$z$};
     \vertex [dot] [right=1.2cm of x] (y) {};
     \vertex [right=0.cm of y] {$0$};
     \vertex [below=0.85cm of x] (v1);
     \vertex [below=.8cm of v1] (v2);
     \vertex [below=0.85cm of v2] (p) {};
     \vertex [below=0.85cm of y] (v1prime); 
     \vertex [below=0.8cm of v1prime] (v2prime); 
     \vertex [below=0.85 of v2prime] (pprime) {};
     \vertex [below right=3.0cm and 0.3cm of x] {(b)};
       \propag [revmom={[arrow shorten=0.1]$p$}] (x) to (v1); 
       \propag (v1) [half left,looseness=1.75, mom={[arrow shorten=0.25, arrow distance=1.5mm] $\ell$}] to (v2);
       \propag (v2) [half left,looseness=1.75, mom={[arrow shorten=0.25, arrow distance=1.5mm] $p+\ell$}] to (v1);
       \propag [revmom={[arrow shorten=0.1]$p$}] (v2) to (p); 
       \propag [revmom'={[arrow shorten=0.3]$p$}] (pprime) to (y);
     \end{feynhand}
   \end{tikzpicture}
 \hspace{1.0truecm}
   \begin{tikzpicture} 
     \begin{feynhand} 
      \vertex [dot] (x) {};
      \vertex [left=0.cm of x] {$z$};
      \vertex [dot] [right=1.2cm of x] (y) {};
      \vertex [right=0.cm of y] {$0$};
      \vertex [below=1.25cm of x] (v1);
      \vertex [below=1.25cm of v1] (p) {};
      \vertex [below=1.25cm of y] (v1prime); 
      \vertex [below=1.25cm of v1prime] (pprime) {};
      \vertex [below right=3.0cm and 0.3cm of x] {(c)};
      \propag [mom'={[arrow shorten=0.2]$\ell$}] (x) to (v1);
      \propag [revmom'={[arrow shorten=0.2]$p$}] (v1) to (p);
      \propag (y) to (v1prime);
      \propag [mom={[arrow shorten=0.2]$p$}] (v1prime) to (pprime);
      \propag [mom={[arrow shorten=.2] $p+\ell$}] (v1) to (v1prime);
     \end{feynhand}
   \end{tikzpicture} 
	\vspace*{5mm}
	\caption{Feynman diagrams up to one loop for $\langle 0 | T\left[\phi\left(z\right)\phi\left(0\right)\phi\left(z_1\right)\phi\left(z_2\right)\right]|0\rangle$.}
	\label{fig:2}
\end{figure}
Denoting the propagator in position space as
\begin{equation}
        \langle 0 | T\left[\phi\left(x\right)\phi\left(y\right)\right]|0\rangle = \wick{\c \phi_x \c \phi_{y} }   \, ,     
\end{equation}
the Wick contraction that contributes to the tree level diagram (a) of Fig.~\ref{fig:2} is given by 
\begin{align}
\label{eq::contraction}
        \wick{\c \phi_z \c \phi_{z_1} \c \phi_{z_2} \c \phi_{0}}=
        \int_{l_1} \frac{i \, e^{-i l_1 \cdot \left(z-z_1\right)}}{l_1^2 -m^2 + i\epsilon} \int_{l_2} 
        \frac{i\,e^{-i l_2 \cdot z_2}}{l_2^2 -m^2 + i\epsilon}\, ,
\end{align}
where we use the notation
\begin{equation}
        \int_k = \int \frac{\mathrm{d}^dk}{(2\pi)^d}\, .
\end{equation}
Plugging Eq.~\eqref{eq::contraction} in Eq.~\eqref{eq::LSZ} we obtain the tree
level expression for $\widehat{\mathcal{M}}\left(\nu, z^2\right)$
\begin{align}
\label{eq::treelevel}
        \widehat{\mathcal{M}}^{(0)}\left(\nu, z^2\right) = - e^{-i \nu} \equiv \widehat{\mathcal{M}}^{(0)}\left(\nu, 0\right) .
\end{align}
Note that the tree level result does not depend on the invariant separation $z^2$ and therefore we can set $z^2=0$ in the second equality above.

At one-loop order the self-energy diagram (b) yields the mass and wave function
renormalization. Its contribution to Eq.~\eqref{eq::LSZ} is  
\begin{align}
        \widehat{\mathcal{M}}_{\text{self}}\left(\nu,z^2\right) 
        &=  R\, \widehat{\mathcal{M}}^{(0)}\left(\nu, 0\right)\, ,
\end{align}
where $R$ is the $\mathcal{O}\left(g^2\right)$ contribution to the residue of the propagator at the pole mass. In $d = 6 -2\epsilon$ dimensions, we have
\begin{align}
        R = {\frac{d\Pi\left(l^2\right)}{dl^2}}_{l^2=p^2_{\text{pole}}} 
        = \alpha\left[\frac{1}{12}\log\frac{m^2}{\mu^2} + \frac{1}{12}\frac{1}{\epsilon} + \frac{b}{2}\right],
\end{align}
where $b/2$ is a finite contribution and $\alpha=g^2 /(64\pi^3) $. The same
$\mathcal{O}\left(\alpha\right)$ contribution is obtained from the diagram with
the self energy corrections on the second leg, so that the total contribution
coming from the tree level plus self-energy corrections is 
\begin{align}\label{eq::zmzphi}
        \widehat{\mathcal{M}}_{\text{self}}\left(\nu,z^2\right)=\left[1+ \alpha\left(\frac{1}{6}\log\frac{m^2}{\mu^2} 
        + \frac{1}{6}\frac{1}{\epsilon} + b\right) \right]\widehat{\mathcal{M}}^{(0)}\left(\nu, 0\right)
        + \mathcal{O}\left(\alpha^2\right)\, .
\end{align}
Note the absence of any $z^2$ dependence: as far as the first two diagrams of
Fig.~\ref{fig:2} are concerned, there are no differences between the light-cone
and the pure spatial case. This is to be expected, since the one-loop diagrams
(b) simply implement the mass and wave function renormalization.

We can now move to the computation of the remaining
$\mathcal{O}\left(\alpha\right)$ term, \ie\ diagram (c). This contraction is
given by
\begin{align}
\int &dw_1\, dw_2\,\wick{\c \phi_z \c \phi_{w_1} \c \phi_{w_1} \c \phi_{z_1} \c \phi_{w_2} \c \phi_{w_1} \c \phi_{w_2} \c \phi_{0} \c \phi_{z_2} \c \phi_{w_2} } = \nonumber\\
&=\left(-ig\right)^2
\int dw_1\,dw_2\,
\int_{l_1} \frac{ie^{-il_1 \cdot \left(z-w_1\right)}}{l_1^2-m^2 + i\epsilon}
\int_{l_2} \frac{ie^{-il_2 \cdot \left(w_1-z_1\right)}}{l_2^2-m^2 + i\epsilon}
\int_{l_3} \frac{ie^{-il_3 \cdot \left(w_2-w_1\right)}}{l_3^2-m^2 + i\epsilon} 
\times \nonumber\\
&\,\,\,\,\,\,\,\,\,\,\,\,\,\,\,\,\,\,\,\,\,\,\,\,\,\,\,\,\,\,\,\,\,\,\,\,\,\,\,\,\,\,\,\,\,\,\,\,\,\,\,\,\,\,\,\,\,\,\,\,\,\,\,\,\,\,\,\,\,\,\,\,\,\,\,\,\,\,\,\, \times
\int_{l_4} \frac{ie^{-il_4 \cdot w_2}}{l_4^2-m^2 + i\epsilon} 
\int_{l_5} \frac{ie^{-il_5 \cdot \left(z_2-w_2\right)}}{l_5^2-m^2 + i\epsilon}.
\end{align}
Plugging this into Eq.~\eqref{eq::LSZ}, we have
\begin{align}
        \label{eq::M1}
        \widehat{\mathcal{M}}^{(1)}\left(\nu, z^2\right) 
        &= -i \,g^2 \int_k 
        \frac{e^{-i k\cdot z}}{\left(k^2-m^2 + i\epsilon\right)^2}
        \frac{1}{\left(p-k\right)^2-m^2 + i\epsilon} \nonumber \\
        & = g^2 \int_0^1 d\xi\,
        \left(1-\xi\right) K\left(z^2, M^2\right)\, 
        \widehat{\mathcal{M}}^{(0)}\left(\xi\nu, 0\right)\, , 
\end{align}
where we have introduced a Feynman parameter $\xi$ and defined 
\begin{align}
\label{eq::kerneldef}
        K\left(z^2, M^2\right) = 2i\int_q
        \frac{e^{-i q\cdot z}}{\left(q^2-M^2 + i\epsilon\right)^3},
\end{align}
with
\begin{align}
& q = k-\xi p\, , \\
& M^2 = m^2\left(1-\xi+\xi^2\right)\, .
\end{align}

The integral $K\left(z^2, M^2\right)$ can be computed by performing a Wick rotation $z_E^\mu = \left(iz^0,\vec{z}\right)$
and using
\begin{align}
        \frac{1}{\left(q_E^2+m^2\right)^{\alpha}} = 
        \frac{1}{\Gamma\left(\alpha\right)}\, 
        \int_0^{\infty} dT \, T^{\alpha-1} e^{-T\left(q_E^2+m^2\right)}\, .
\end{align}
We obtain
\begin{align}
\label{eq::kernel}
        K\left(z^2, M^2\right) 
        &= 2\int \frac{d^dq_E}{\left(2\pi\right)^d}\, 
        \frac{e^{i q_E z_E}}{\left(q_E^2+M^2\right)^3} 
        = \int_0^{\infty} dT\, T^2 e^{-T M^2} 
        \int \frac{d^dq_E}{\left(2\pi\right)^d}\, 
        e^{i q_E z_E -T q_E^2} \nonumber\\ \nonumber\\
        &= \frac{1}{\left(4\pi\right)^\frac{d}{2}} 
        \int_0^{\infty}\frac{dT}{T}T^{3-\frac{d}{2}}\, 
        e^{-TM^2} e^{-\frac{z_E^2}{4T}}\, ,
\end{align}
where in the last line we have performed the Gaussian integral over $d^{d}q_E$.

Since we are considering the case of a light-cone separation $z_E^2 = -z^2 =0$,
$K\left(0,M^2\right)$ in $d =6$ dimensions is logarithmically divergent. The
divergence arises from the lower end of the integral over $T$, as the
exponential suppression factor in the integrand vanishes on the light-cone. We
apply dimensional regularization, taking $d = 6 -2\epsilon$ and
introducing the $\overline{\mathrm{MS}}$ scale $\mu$ through the rescaling of
the coupling $g^2\rightarrow g^2 e^{\gamma_E}\mu^2/(4\pi)$. We find
\begin{align}
        K\left(0, M^2; \mu^2\right) 
        &= \int_0^{\infty} \frac{dT}{T}\, 
        \left(T\mu^2e^{\gamma_E}\right)^{\epsilon}e^{-TM^2} 
        = \Gamma\left(\epsilon\right) 
        \left(\frac{\mu^2e^{\gamma_E}}{M^2}\right)^{\epsilon} 
        =\frac{1}{\epsilon} + \log\frac{\mu^2}{M^2}\, ,
\end{align}
where the pole in $1/\epsilon$ reflects the original logarithmic divergence in
dimensional regularization. Putting everything together, we obtain the full
one-loop expression of the bare position space matrix element in dimensional
regularization
\begin{align}
\label{eq::Ioffe1loop}
        \widehat{\mathcal{M}}\left(\nu, 0\right) 
        =& \left[1+ \alpha\left(\frac{1}{6}\log\frac{m^2}{\mu^2} 
        + \frac{1}{6}\frac{1}{\epsilon} + b\right)\right]\, 
        \widehat{\mathcal{M}}^{(0)}\left(\nu, 0\right)
        \nonumber \\ 
        & + \alpha \int_0^1 d\xi\,
        \left(1-\xi\right) \left(\frac{1}{\epsilon} + 
        \log\frac{\mu^2}{m^2\left(1-\xi+\xi^2\right)}\right)\, 
        \widehat{\mathcal{M}}^{(0)}\left(\xi \nu, 0\right)\, .
\end{align}
The structure of the divergences in Eq.~\eqref{eq::Ioffe1loop} shows that this
quantity can be renormalized by convolution with a renormalization kernel
$\mathcal{K}$. Denoting the renormalized matrix element as
$\widehat{\mathcal{M}}_R\left(\nu,0,\mu^2\right)$, we have
\begin{align}
        \widehat{\mathcal{M}}_R\left(\nu,0,\mu^2\right) = \int_0^1 dy\, 
        \mathcal{K}\left(y\right)\, 
        \widehat{\mathcal{M}}\left(y\nu, 0\right)\, .
\end{align}
The specific choice of the finite terms that appear in the kernel
$\mathcal{K}\left(y\right)$, together with subtraction of the $1/\epsilon$
poles, defines the renormalization scheme. For example, in the
$\overline{\mathrm{MS}}$ scheme, the renormalization kernel is
\begin{align}
        \label{eq:renormalizationkernel}
        \mathcal{K}\left(y\right) = 
        \delta\left(1-y\right) - 
        \alpha\, \left[ \frac{1}{6\,\epsilon}\delta\left(1-y\right) 
        + \frac{1}{\epsilon}\left(1-y\right)\right]\, ,
\end{align}
and the corresponding renormalized quantity is
\begin{align}
        \label{eq::Ioffe1loopren}
        \widehat{\mathcal{M}}_R\left(\nu,0,\mu^2\right) 
        =& \left[1 + 
        \alpha \left(\frac{1}{6}\log\frac{m^2}{\mu^2} + b\right)\right]\, 
        \widehat{\mathcal{M}}^{(0)}\left(\nu, 0\right) \nonumber\\ 
        & + \alpha\,  \int_0^1 d\xi\,\left(1-\xi\right)
        \log\frac{\mu^2}{m^2\left(1-\xi+\xi^2\right)}\, 
        \widehat{\mathcal{M}}^{(0)}\left(\xi\nu, 0\right)\, .  
\end{align}
We conclude this derivation with a comment on the form of the renormalization
kernel $\mathcal{K}$ given in Eq.~\eqref{eq:renormalizationkernel}: the
contribution proportional to a delta function is a multiplicative
renormalization term, implementing the subtraction of the singularities
generated by diagram (b) of Fig.~\ref{fig:2}, which is basically the wave
function renormalization. The second contribution,
$-\frac{\alpha}{\epsilon}\left(1-y\right)$, implements the renormalization of
the one-loop diagram (c) of Fig.~\ref{fig:2}, and because this contribution is
not proportional to a delta function, the renormalization of this term is not
multiplicative, but requires a convolution.

Taking the log derivative of Eq.~\eqref{eq::Ioffe1loopren} we obtain
\begin{align}
        \label{eq:DGLAPpositionspace}
        \mu^2\frac{d}{d\mu^2}\, \widehat{\mathcal{M}}_R\left(\nu,0,\mu^2\right) 
        = \alpha\int_0^1 d\xi\,P\left(\xi\right)\,\widehat{\mathcal{M}}_R\left(\xi\nu,0,\mu^2\right) 
        + \mathcal{O}\left(\alpha^2\right)\, ,
\end{align}
where the $\mathcal{O}\left(\alpha\right)$ Altarelli-Parisi splitting kernel is given by
\begin{align}
        \label{eq:AltarelliParisikernel}
        P\left(\xi\right) = 
        \left(1-\xi\right) - \frac{1}{6}\delta\left(1-\xi\right) = 
        \left(1-\xi\right)_+ + \frac{1}{3}\delta\left(1-\xi\right)\, ,
\end{align}
with the action of the plus distribution over a generic test function $g\left(\xi\right)$ defined as
\begin{align}
	\label{eq:plus}
	\int_0^1 dx\,\left(1-\xi\right)_+g\left(\xi\right) = \int_0^1 dx\,\left(1-\xi\right)\left[g\left(\xi\right) - g\left(1\right)\right].
\end{align}
The renormalized collinear PDF is defined from the renormalized matrix element,
\begin{align}
        \label{eq::PDFs}
        \widehat{\mathcal{M}}_R\left(\nu,0,\mu^2\right) = 
        \int_{0}^{1} dx \, e^{ix\nu}\widehat{f}\left(x,\mu^2\right)\, ,
\end{align}
and therefore, from Eq.~\eqref{eq:DGLAPpositionspace},
\begin{align}
        \label{eq:AP}
        \mu^2\frac{d}{d\mu^2}\, \widehat{f}\left(x, \mu^2\right) = 
        \alpha \int_{x}^{1} \frac{d\xi}{\xi}\,P\left(\xi\right)\,\widehat{f}\left(\frac{x}{\xi}, \mu^2\right)\, ,
\end{align}
which yields the standard DGLAP evolution equations, which were already obtained
in Ref.~\cite{Collins:1980ui} for the scalar theory. The solution of
Eq.~\eqref{eq:AP} in perturbation theory is given by an evolution kernel
$\Gamma\left(x,\mu,\mu_0,\alpha\right)$, which allows the PDF at a generic scale
$\mu$ to be computed in terms of the PDF at the scale $\mu_0$ as
\begin{align}
        \label{eq:solutionDGLAP}
	\widehat{f}\left(x,\mu^2; \theta\right) = \int_x^1 \frac{d\xi}{\xi}\,\Gamma\left(\frac{x}{\xi},\mu,\mu_0,\alpha_s\right)
	\widehat{f}\left(\xi,\mu_0^2; \theta\right)\, .
\end{align}

\section{Spatial separation}
\label{sec:spatial-separtaion}

We now consider the case in which the separation between the fields is purely
spatial $z_E^2 = z_3^2$. As seen in the previous section, the $z^2$ dependence
enters only through diagram (c) of Fig.~\ref{fig:2}. Considering this
contribution, the kernel $K\left(z^2,M^2\right) $ defined in
Eq.~\eqref{eq::kernel} is no longer divergent for $z_3\neq 0$, as the term
$\exp\left[-z_E^2/(4T)\right]$ regulates the small-$T$ behaviour. The integral
can evaluated directly in $d=6$ dimensions, yielding
\begin{align}
    \label{eq::Bessel}
    K\left(-z_3^2, M^2\right) = 
    \frac{1}{64\pi^3}\int_0^{\infty}\frac{dT}{T} e^{-T} e^{-\frac{\left(M z_3\right)^2}{4T}} =\frac{1}{64\pi^3}\, 2K_0\left(M z_3\right),
\end{align}
where $K_0$ is the modified Bessel function. Plugging Eq.~\eqref{eq::Bessel} into Eq.~\eqref{eq::M1} we obtain the contribution from diagram (c) in the case of purely spatial separation:
\begin{align}
\label{eq::1loopcont}
    \widehat{\mathcal{M}}^{(1)}\left(\nu, -z_3^2\right) = 
    \alpha\int_0^{1} d\xi \, \left(1-\xi\right) 2K_0\left(M z_3\right)\widehat{\mathcal{M}}^{(0)}\left(\xi\nu, 0\right) .
\end{align}
Note that, as long as $z_3 \neq 0$, this contribution does not contain any UV
divergences. For $M z_3 \rightarrow 0$ the Bessel function diverges
logarithmically, and we recover the UV divergence of the light-cone case.

The full one-loop bare matrix element is then given by
\begin{align}
\label{eq::pIoffe1loop}
    \widehat{\mathcal{M}}\left(\nu, -z_3^2\right) = & \left[1+ \alpha\left(\frac{1}{6}\log\frac{m^2}{\mu^2} + \frac{1}{6}\frac{1}{\epsilon} + b\right)  \right]\widehat{\mathcal{M}}^{(0)}\left(\nu, 0\right) \nonumber \\
    &+\alpha \int_0^1 d\xi\,\left(1-\xi\right)2K_0\left(M z_3\right)  \widehat{\mathcal{M}}^{(0)}\left(x \nu, 0\right).
\end{align}
As before, this quantity can be renormalized by convolution, 
\begin{align}
    \widehat{\mathcal{M}}_R\left(\nu, -z_3^2;\,\mu^2\right)= 
    \int_0^1 dy \, \tilde{\mathcal{K}}\left(y\right) \widehat{\mathcal{M}}\left(y\nu, -z_3^2\right)\, .
\end{align}
However, since the only UV pole comes from the self-energy contributions, the
kernel $\tilde{\mathcal{K}}\left(y\right)$ is proportional to a delta function.
For example, in the \msbar\ scheme we can take
\begin{align}
    \tilde{\mathcal{K}}\left(y\right) = 
    \delta\left(1-y\right)\left[ 1-\alpha \frac{1}{6\,\epsilon} \right].
\end{align}
In other words, in the case of purely spatial separation the renormalization of
the matrix element is purely multiplicative \cite{Ishikawa:2017faj}. The
additional UV divergence we had to remove in the light-cone case is substituted
here by a finite contribution $K_0\left(M z_3\right) $. The corresponding
renormalized quantity is
\begin{align}
\label{eq::pIoffe1loopren}
    \widehat{\mathcal{M}}_R\left(\nu, -z_3^2;\,\mu^2\right) =&
    \left[1+ \alpha\left(\frac{1}{6}\log\frac{m^2}{\mu^2} + b\right)  \right]\widehat{\mathcal{M}}^{(0)}\left(\nu, 0\right) \nonumber \\
    &+\alpha \int_0^1 d\xi\,\left(1-\xi\right)2K_0\left(M z_3\right)  \widehat{\mathcal{M}}^{(0)}\left(x \nu, 0\right).
\end{align}
Note also that both Eqs.~\eqref{eq::Ioffe1loopren} and \eqref{eq::pIoffe1loopren} contain an infrared (IR) divergence regularized by the mass $m$: 
in the former the mass is manifest in the $\log$, while in the latter the mass appears in the Bessel function, which diverges logarithmically for $m\rightarrow 0$.

\section{Factorization theorem}
\label{sec:factorization}

Having defined the renormalized correlators in the previous sections, let us
investigate the one-loop relation between the light-cone and the equal-time
correlators. Combining Eqs.~\eqref{eq::Ioffe1loopren} and
\eqref{eq::pIoffe1loopren} we write
\begin{align}
	\label{eq::fact00}
	\widehat{\mathcal{M}}_R\left(\nu,-z_3^2;\,\mu^2\right) = 
	& {} \widehat{\mathcal{M}}_R\left(\nu,0,\mu^2\right) \nonumber \\
	& + \alpha\int_0^1 d\xi \, 
	\left(1-\xi\right) \left(2K_0\left(M z_3\right)
	-\log\frac{\mu^2}{M^2}\right)\, 
	\widehat{\mathcal{M}}_R\left(\xi\nu,0,\mu^2\right)\, ,
\end{align}
and using Eq.~\eqref{eq::PDFs} we find
\begin{align}
	\label{eq::fact0}
	\widehat{\mathcal{M}}_R\left(\nu, -z_3^2; \mu^2\right) = 
	\int_{0}^{1} dx\,\tilde{C}\left(x\nu, m z_3, \frac{\mu^2}{m^2} \right) \widehat{f}\left(x,\mu^2\right)\, ,
\end{align}
with
\begin{align}
	\tilde{C}\left(x\nu, m z_3, \frac{\mu^2}{m^2} \right) = 
	e^{i x\nu} - \alpha\int_0^1 d\xi \, \left(1-\xi\right)
	\left(2K_0\left(M z_3\right)-\log\frac{\mu^2}{M^2}\right) e^{i \xi x\nu}\, .
\end{align}
This expression shows the connection between the collinear PDFs and an
equal-time correlator, through a convolution with a perturbative kernel. In
general, the latter contains a logarithmic dependence on $m^2$, namely the
kernel contains IR singularities. However, as we will see, these
singularities cancel exactly when taking a specific limit, leaving an expression
free from IR poles, which therefore has   the form of a proper factorization
theorem.
Before discussing this in detail, we recall that, although Eq.~\eqref{eq::fact0}
has been worked out in perturbation theory, considering matrix elements between
on-shell quark states, the renormalization of the bilocal operators discussed so
far does not depend on our choice of specific external states. It follows that
Eq.~\eqref{eq::fact0} holds also for external proton states. From now on we will
refer to full proton matrix elements rather than partonic ones, removing the
symbol `$\,\,\widehat{}\,\,$' used so far to denote partonic quantities.

\subsection{Factorization theorem in position space: small-$z_3^2$ limit}

The behavior of the coefficient $\tilde{C}$ in the small-$z_3^2$ limit is
obtained by expanding the Bessel function as 
\begin{align}
	\label{eq::besselexpansion}
	2K_0\left(M z_3\right) = - \log\left(M^2 z_3^2\right) + 2 \log\left(2e^{-\gamma_E}\right) + \mathcal{O}\left(M^2 z_3^2\right)\, ,
\end{align} 
so that Eq.~\eqref{eq::fact0} becomes
\begin{align}
	\label{eq::fact2}
	\mathcal{M}_R\left(\nu, -z_3^2; \, \mu^2\right) &=  
	\int_{0}^{1} dx \, \tilde{C}\left(x\nu,\mu^2 z_3^2 \right) 
	f\left(x,\mu^2\right)\, ,
\end{align}
with
\begin{align}
	\label{eq::Cpseudo0}
	\tilde{C}\left(x\nu,\mu^2 z_3^2 \right) =
	 e^{i x\nu} - \alpha\int_0^1 d\xi \, 
	 \left(1-\xi\right) 
	 \log\left( \mu^2 z_3^2\frac{e^{2\gamma_E}}{4} \right) e^{i \xi x\nu}
	 + \mathcal{O}\left(m^2 z_3^2\right)\, .
\end{align}
We note that in this limit the logarithmic behaviour of the Bessel function
matches that of the light-cone quantity, so that the two matrix elements display
the same IR behaviour: as a result the coefficient $\tilde{C}$ is IR safe, and
Eq.~\eqref{eq::fact2} represents a proper factorization theorem connecting a
lattice computable quantity on the left hand side with a collinear PDF on the
right hand side. 

We note that this factorization also applies to the so-called reduced
distributions \cite{Zhang:2018ggy,Radyushkin:2018cvn}, the quantities usually
determined in lattice calculations in the pseudo-PDF approach, first introduced
in Ref.~\cite{Radyushkin:2017cyf}. They were originally defined as
\begin{align}
\label{eq::reduced}
	\mathfrak{M}\left(\nu,-z_3^2\right) =
	\frac{\mathcal{M}_R\left(\nu, -z_3^2; \mu^2\right)}{\mathcal{M}_R\left(0, -z_3^2; \mu^2\right)}\,,
\end{align}
although a double ratio was proposed in \cite{Orginos:2017kos}. Here we restrict our attention to the ratio defined in Eq.~\eqref{eq::reduced}.
In the context of our model, using the small-$z_3^2$ limit of Eq.~\eqref{eq::fact00} we have
\begin{align}
	\mathfrak{M}\left(\nu,-z_3^2\right) &= 
	\mathcal{M}_R\left(\nu,0, \mu^2\right)  
	-\alpha \log\left( \mu^2 z_3^2\frac{e^{2\gamma_E}}{4} \right) \int_0^1 d\xi\,\left(1-\xi\right)  
	\left[\mathcal{M}_R\left(\xi \nu,0,\mu^2\right)- \mathcal{M}_R\left(\nu,0,\mu^2\right) \right] \nonumber \\
	& = \mathcal{M}_R\left(\nu, 0, \mu^2\right)  
	-\alpha \log\left( \mu^2 z_3^2\frac{e^{2\gamma_E}}{4} \right) \int_0^1 d\xi\,\left(1-\xi\right)_+ 
	\mathcal{M}_R\left(\xi \nu,0, \mu^2\right)\, ,
\end{align}
and therefore
\begin{align}
	\label{eq::factReduced}
	\mathfrak{M}\left(\nu,-z_3^2\right) &=  
	\int_{0}^{1} dx \, \tilde{C}_+\left(x\nu,\mu^2 z_3^2 \right) f\left(x,\mu^2\right)\,,
\end{align}
with
\begin{align}
	\label{eq::Cpseudo}
	\tilde{C}_+\left(x\nu,\mu^2 z_3^2 \right) =
	 e^{i x\nu} - \alpha\,\log\left( \mu^2 z_3^2\frac{e^{2\gamma_E}}{4} \right)
	 \int_0^1 d\xi \, \left(1-\xi\right)_+ e^{i \xi x\nu} + \mathcal{O}\left(m^2 z_3^2\right)\, .
\end{align}
Note the absence of any $\mu^2$ dependence on the left hand side of
Eqs.~\eqref{eq::reduced} and \eqref{eq::factReduced}: the perturbative
dependence on the renormalization scale $\mu^2$ cancels exactly in the ratio,
leaving a quantity that depends only on the scale $z_3^2$. More precisely,
Eqs.~\eqref{eq::factReduced}, \eqref{eq::Cpseudo} show how, in the small-$z_3^2$
limit, the renormalization scale dependence of
$\mathcal{M}_R\left(\nu,0,\mu^2\right)$ generated by diagram (c) is replaced by
an equal $z_3^2$ dependence that can be obtained from the former through the
substitution  
$$\mu^2 \rightarrow \frac{4 e^{-2\gamma_E}}{z_3^2}\, .$$ In other words, the
factorization formula worked out in this section predicts a logarithmic
dependence on $z_3^2$ for the equal-time correlator, which replaces the
analogous logarithmic behaviour of the PDFs on the renormalization scale
$\mu^2$, predicted by the one-loop DGLAP. Such dependence on $z_3^2$ should be
visible in real lattice QCD data when working in the factorization regime, and
indeed it was observed in Refs.~\cite{Radyushkin:2018cvn, Orginos:2017kos}.

\subsection{Factorization theorem in momentum space: large $P_3$ limit}
\label{sec::momentumspace}

A factorization theorem can also be established working in momentum rather than
in position space. Taking the Fourier transform of Eq.~\eqref{eq::fact2} with
respect to $z_3$ and defining
\begin{align}
	\label{eq::qpdf}
	&q\left(y, \mu^2, P_3^2\right) = 
	\frac{P_3}{2\pi} 
	\int_{-\infty}^{\infty}dz_3\, e^{-i y P_3 z_3} 
	\widehat{\mathcal{M}}\left(P_3 z_3, -z_3^2\right)\, , \\
	\label{eq::matching0}
	&C\left(\eta,\frac{m^2}{x^2 P_3^2}, \frac{\mu^2}{m^2}\right) = 
	\int_{-\infty}^{\infty}\frac{d\theta}{2\pi}\, e^{-i\theta\eta}\,
	\tilde{C}\left(\theta, \frac{m\theta}{x P_3}, \frac{\mu^2}{m^2} \right)\, ,
\end{align}
we obtain
\begin{align}
	\label{eq::matching1}
	q\left(y, \mu^2, P_3^2\right) = 
	\int_{0}^{1} \frac{d x}{x}\, f\left(x,\mu^2\right) 
	C\left(\frac{y}{x},\frac{m^2}{x^2 P_3^2}, \frac{\mu^2}{m^2}\right)\, ,
\end{align}
with
\begin{align}
	\label{eq::C0}
	C\left(\eta,\frac{m^2}{x^2 P_3^2}, \frac{\mu^2}{m^2}\right) = 
	\int_0^1 d\xi \, \left(1-\xi\right) 
	\left[\frac{1}{\sqrt{\left(\eta-\xi\right)^2
	 + \frac{M^2}{x^2P_3^2}}} - 
	 \delta\left(\xi-\eta\right) \log\frac{\mu^2}{M^2}\right]\, .
\end{align}
Note that taking the Fourier transform, as in Eq.~\eqref{eq::qpdf}, involves an
integration of the Bessel function $K_0\left(z_3 M\right)$ through its
singularity at $z_3=0$, which is discussed in detail in App.~\ref{app:plus}.
Looking at Eq.~\eqref{eq::C0}, we note that, again, the coefficient $C$ contains
explicit logarithms of the mass, rendering it infrared divergent. However, these
divergences cancel when considering the large $P_3$ regime, by expanding the
Fourier transform of the Bessel function in the limit
$\frac{M^2}{\xi^2P_3^2}\rightarrow 0$. If $\eta>1$ or $\eta<0$, then looking at
Eq.~\eqref{eq::C0} we have 
\begin{align}
	\lim_{P_3\rightarrow \infty} 
	C\left(\eta,\frac{m^2}{x^2 P_3^2}, \frac{\mu^2}{m^2}\right) &= 
	C\left(\eta\right) = 
	\pm \int_0^1 d\xi \,\frac{1-\xi}{\eta-\xi} = 
	\pm \left[\left(1-\eta\right) \log\frac{\eta}{\eta-1} + 1 \right]\, ,
\end{align}
where the solution with the plus refers to $\eta>1$, and the one with the minus
to $\eta<0$. On the other hand, if $\eta \in \left(0,1\right)$, the factor
$1/|\eta-x|$ generated in this limit produces a non-integrable singularity at
$\eta = x$~\cite{Radyushkin:2017lvu}. To overcome this issue, as detailed in App.~\ref{app:plus}, we can
write
\begin{align}
	\label{eq::limit} 
	\frac{1}{\sqrt{\left(\eta-\xi\right)^2 + \frac{M^2}{x^2 P_3^2}}} 
	= \log 4\eta\left(1-\eta\right) \frac{x^2P_3^2}{M^2}
	\delta\left(\eta-\xi\right) + \frac{1}{|\eta-\xi|_+} + \mathcal{O}\left(\frac{M^2}{P_3^2}\right),
\end{align}
so that in the region $\eta \in \left(0,1\right)$  we find
\begin{align}
	C&\left(\eta,\frac{M^2}{x^2 P_3^2}, \frac{\mu^2}{M^2}\right) 
	 \,\,
	\stackrel{P_3\rightarrow \infty}{\sim}\,\, C\left(\eta,\frac{\mu^2}{x^2 P_3^2}\right) \nonumber\\
	&=\int_0^1 d\xi\, \left(1-\xi\right) 
	\left[\frac{1}{|\eta-\xi|_+} + \delta\left(\eta-\xi\right)
	\log 4\eta\left(1-\eta\right)
	\frac{x^2P_3^2}{\mu^2} \right]  + \mathcal{O}\left(\frac{m^2}{P_3^2}\right) \nonumber \\
	&=2\eta -1 + \left(1-\eta\right)
	\log 4\eta\left(1-\eta\right) \frac{x^2P_3^2}{\mu^2} + \mathcal{O}\left(\frac{m^2}{P_3^2}\right). 
\end{align}
Note the cancellation of the logarithmic dependence on the mass, which leads
again to a proper factorization formula, this time in momentum space. We
conclude that, in momentum space, the factorization theorem is realized in the
limit $P_3\rightarrow \infty$ and in our model this factorization theorem takes
the form
\begin{align}
	\label{eq::factmomentum}
	q\left(y, \mu^2, P_3^2\right) = 
	\int_{0}^{1} \frac{dx}{x}\, 
	f\left(x,\mu^2\right) 
	C\left(\frac{y}{x},\frac{\mu^2}{x^2 P_3^2}\right) + \mathcal{O}\left(\frac{m^2}{P_3^2}\right)\, ,
\end{align}
with
\begin{equation}
	\label{eq::matching}
	\begin{split}
	C\left(\eta,\frac{\mu^2}{x^2 P_3^2} \right)&= \delta\left(1-\eta\right) + \alpha 
	\begin{cases} 
	\left(1-\eta\right)\log\frac{\eta}{\eta-1} + 1 
	\,\,\,\,\,\,\,\,\,\,\,\,\,\,\,\,\,\,\,\,\,\,\,\,\,\,\,\,\,\,\,\,\,\,\,\,\,\,\,\,\,\,\,\,\,\,\,\,\,\,\, \eta > 1\\ 
	\left(1-\eta\right)\log 4\eta\left(1-\eta\right)\frac{x^2P_3^2}{\mu^2} + 2\eta -1 \,\,\,\,\,\,\,\,\,\, 0<\eta < 1 \\ 
	-\left(1-\eta\right)\log\frac{\eta}{\eta-1} - 1 \,\,\,\,\,\,\,\,\,\,\,\,\,\,\,\,\,\,\,\,\,\,\,\,\,\,\,\,\,\,\,\,\,\,\,\,\,\,\,\,\,\,\,\,\,\, \eta<0
	\end{cases}
	\,.
	\end{split}
\end{equation}

Factorization in position space, given in Eqs.~\eqref{eq::fact2} and
\eqref{eq::Cpseudo0}, is equivalent to factorization in momentum space, given in
Eqs.~\eqref{eq::factmomentum} and \eqref{eq::matching}. In other words, taking
the small-$z_3^2$ limit in position space is entirely equivalent to taking the
large-$P_3$ limit in momentum space. This can be easily verified by computing
the Fourier transfom of the small-$z_3^2$ coefficient $\tilde{C}$ of
Eq.~\eqref{eq::Cpseudo0}, and checking that it is equal to the high-$P_3$
coefficient $C$ of Eq.~\eqref{eq::matching} 
\begin{align}
	\label{eq::check}
	\frac{1}{x}\,C\left(\eta,\frac{\mu^2}{x^2 P_3^2}\right) &= 
	\frac{P_3}{2\pi}\int_{-\infty}^{\infty} dz_3\, e^{-i y P_3 z_3 }\,
	\tilde{C}\left(x\nu, \mu^2 z_3^2 \right) \nonumber \\
	&= \frac{1}{x}\,\int_{-\infty}^{\infty}\frac{d\theta}{2\pi}\, e^{-i\theta\eta}\,
	\tilde{C}\left(\theta, \frac{\mu^2\theta^2}{x^2 P_3^2} \right)\,\,\,\,\, \text{with}\,\,\,\, \eta = \frac{y}{x}.
\end{align}

This check, despite being conceptually straightforward, does require some
care~\cite{Izubuchi:2018srq}. We provide the details of the computation in
App.~\ref{app:plus}.
The implementation of the factorization theorem in position space, together with
the definition of reduced distributions, are the typical approach followed in nonperturbative calculations of
pseudo-PDFs~\cite{Radyushkin:2017cyf,Orginos:2017kos,Joo:2019jct,Joo:2019bzr,Joo:2020spy,
Radyushkin:2019owq}, while the realization of the factorization in momentum
space characterizes the quasi-PDF
approach~\cite{PhysRevLett.110.262002,Alexandrou:2018pbm, Alexandrou:2019lfo,
Chai:2020nxw, Bhat:2020ktg}. 

As we have shown in this section in the simplified context of our model, these
two approaches are conceptually equivalent, and related by a Fourier transform:
in one case the lattice calculation needs to provide the correlators for  small
values of $z_3$, while in the other large values of $P_3$ are required. In both
scenarios, however, the object that is actually computed is the matrix element
of spatially-separated fields. This is the only quantity of interest, without
the need to define either pseudo- or quasi-PDFs.

\section{Smeared distributions}
\label{sec:flow}

In Ref.~\cite{Monahan:2016bvm,Monahan:2017hpu}, the gradient flow was proposed
as an approach to control the power divergence associated with the Wilson-line
operator that defines the Ioffe time distribution in QCD. The gradient
flow~\cite{Narayanan:2006rf,Luscher:2011bx,Luscher:2013cpa} is a classical,
gauge-invariant, one-parameter mapping of the theory that exponentially damps
the UV fluctuations. This corresponds to smearing in real space, with a smearing
scale that is parametrised by the flow time. In the limit of small flow time,
the matrix elements of smeared fields can be related to those at vanishing flow
time by a short flow-time expansion~\cite{Luscher:2013vga}.

In Yang-Mills theories, gauge invariance ensures that no new divergences are
introduced at finite flow time. Thus, provided the boundary theory is properly
renormalized, the matrix elements of composite operators composed of fields at
finite flow time are guaranteed to be finite. In the absence of gauge
symmetries, the simplest method for maintaining this property is to exclude
interactions from the flow time evolution of the fields, in which case this
evolution corresponds to simple Gaussian
smearing~\cite{Monahan:2015lha,Monahan:2015fjf,Fujikawa:2016qis}. 

The flow time can be viewed as a non-perturbative regulator that does not affect
the infrared properties of correlation functions. The smeared Ioffe-time matrix
elements, constructed from fields at finite flow time, therefore satisfy the
same factorization theorems as the original Ioffe-time matrix
elements~\cite{Monahan:2016bvm}. In the scalar case, the boundary fields
$\phi(x)$ in Eq.~\eqref{eq::MEpartonic} are replaced by fields at finite
flow time $\rho(t;x)$, so that the partonic matrix element becomes
\begin{align}
        \label{eq::MEflow}
        \widehat{\mathcal{M}}_t\left(\nu,\overline{z}^2\right) = 
        \langle p | \rho\left(t;z\right)\rho\left(t;0\right)  | p \rangle\, .
\end{align}
Here the subscript indicates that the fields are evaluated at flow time $t$, and
$\overline{z}^2 = z^2/t$. 

The gradient flow is only well-defined in Euclidean space, but for $z^2 <0$, the
matrix elements are signature independent~\cite{Briceno:2017cpo}. The tree-level
and one-loop diagrams that contribute to this matrix element are exactly those
given in Fig.~\ref{fig:2}, with $\phi(x)$ replaced by $\rho(t;x)$. Working in
the small flow-time regime, where contributions of ${\cal O}(t)$ can be
neglected, the only diagram that must be calculated is diagram (c) of
Fig.~\ref{fig:2}. Therefore, we can deduce the factorization properties of the
smeared matrix element directly from the analogue of Eq.~\eqref{eq::M1} at
nonzero flow time
\begin{align}
   \label{eq::M1t}
   \widehat{\mathcal{M}}_t^{(1)}\left(\nu, -\overline{z}_3^2\right) 
   &= g^2 \int_{k_E}  e^{-2k_E^2t}\,
   \frac{e^{-i k_{\mathrm{E}}z_3}}
        {\left(k_{\mathrm{E}}^2+m^2\right)^2}
   \frac{1}{\left(p_{\mathrm{E}}-k_{\mathrm{E}}\right)^2+m^2} \nonumber \\
   &= g^2 \int_0^1 d\xi\,
        \left(1-\xi\right) K_t\left(-\overline{z}_3^2, \overline{M}^2\right)\, 
        \widehat{\mathcal{M}}^{(0)}\left(\xi\nu, 0\right)\, ,
\end{align}
where the exponential damping is the result of the smearing of the fields and we
have introduced $\overline{M}^2 = M^2 t$. Here the kernel
$K_t\left(-\overline{z}_3^2, \overline{M}^2\right)$ is given by
\begin{equation} 
        \label{eq:ktd}
        K_t\left(-\overline{z}_3^2, \overline{M}^2\right) =  
        \frac{\mu^{6-d}}{(4\pi)^{d/2}} e^{-2m^2t\xi}\, 
        \int_0^\infty \mathrm{d}T\,
                \frac{T^2}{(T+2t)^{d/2}}\, e^{-TM^2} 
                e^{(4\xi t p_{\mathrm{E}} - iz_{\mathrm{E}})^2/(4(T+2t))}\, ,
\end{equation}
which reduces to the kernel in Eq.~\eqref{eq::kernel} when $t = 0$.

By introducing the further dimensionless variables $\overline{\mu}^2 = \mu^2 t$,
$\overline{m}^2 = m^2 t$, and 
\begin{equation}
        \beta^2 = -\frac{1}{t} 
        \left(\xi t p_{\mathrm{E}}^\mu - \frac{iz_{\mathrm{E}}^\mu}{2}\right)^2
         = \xi^2 \overline{m}^2+i \xi \nu + 
         \frac{\overline{z}_3^2}{4}\, ,
\end{equation}
and changing variables to $u = T/t + 2$, the integral becomes
\begin{equation}
        K_t\left(-\overline{z}_3^2, \overline{M}^2\right) = 
        \frac{\overline{\mu}^{6-d}}{(4\pi)^{d/2}} e^{-2(\xi-1)^2\overline{m}^2}
        \int_{2}^\infty \mathrm{d}u\,
        \frac{(u-2)^2}{u^{d/2}} e^{-u\overline{M}^2-\beta^2/u}\, .
\end{equation}
This integral can be solved in terms of {\em incomplete Bessel
functions}~\cite{Cichetti:2004,Jones:2007,Harris:2008}, which can be studied in
various asymptotic regimes. In particular,
\begin{align}
        \label{eq:Ktdimless}
        K_t\left(-\overline{z}_3^2, \overline{M}^2\right) = 
        {} & \frac{2\,\overline{\mu}^{6-d}}{(4\pi)^{d/2}}
        e^{-2\frac{(\xi-1)^2}{1-\xi+\xi^2}\overline{M}^2} \nonumber \\
        {} & \times \left[K_0(2\,|\overline{M}\beta|,2) - 
        4\frac{\overline{M}}{|\beta|} 
        K_1(2\,|\overline{M}\beta|,2) + 
        4\frac{\overline{M}^2}{\beta^2} K_2(2\,|\overline{M}\beta|,2)\right]\, ,
\end{align}
where
\begin{equation}
        \label{eq:Kndef}
        K_n(y,a) = K_n(y) - J(y,n,a)\, ,
\end{equation}
with $J(y,n,a)$ the finite integral
\begin{equation}
        \label{eq:J}
        J(y,n,a) = \int_0^a \mathrm{d}v\, e^{-y \cosh(v)}\cosh(nv)\, .
\end{equation}

This result is finite in six dimensions, because the incomplete Bessel functions
are finite at finite flow time and quark mass. Indeed, one can evaluate these
integrals numerically by imposing a cutoff. 
For sufficiently large cutoff, the
results are independent of the cutoff value. 
Using Eq.~\eqref{eq:Kndef}, Eq.~\eqref{eq:Ktdimless} can be written as
\begin{align}
        \label{eq:KtBessel}
        K_t\left(-\overline{z}_3^2, \overline{M}^2\right) = 
        {} & \frac{2}{(4\pi)^3} 
        e^{-2\frac{(\xi-1)^2}{1-\xi+\xi^2}\overline{M}^2} 
        \bigg\{
        K_0(2\,|\overline{M}\beta|) - 
        4\frac{\overline{M}}{|\beta|} K_1(2\,|\overline{M}\beta|) + 
        4\frac{\overline{M}^2}{\beta^2} K_2(2\,|\overline{M}\beta|) \nonumber \\
        {} & - J(2\,|\overline{M}\beta|,0,2) +4\frac{\overline{M}}{|\beta|} J(2\,|\overline{M}\beta|,1,2) -4 \frac{\overline{M}^2}{\beta^2}J(2\,|\overline{M}\beta|,2,2)\bigg\}\, .
\end{align}
In the limit where 
\begin{align}
        \frac{t^2 m^2}{z_E^2} \ll 1\, ,
\end{align}
the argument of the Bessel functions, $|\overline{M}\beta|$, can be expressed as
\begin{align}
        2|\overline{M}\beta| = M |z_E| 
                + \mathcal{O}\left(\frac{t^2 m^2}{z_E^2}\right) 
        = M z_3 + \mathcal{O}\left(\frac{t^2 m^2}{z_E^2}\right)\, ,
\end{align}
so that, in the limit of small $z_3$ we can expand them as
\begin{align}
        2K_0(2\,|\overline{M}\beta|) = {} & - \log\left(M^2 z_3^2\right) 
        + 2 \log\left(2e^{-\gamma_E}\right) + 
        \mathcal{O}\left(m^2 z_3^2,\frac{t^2 m^2}{z_E^2}\right)\,,
        \label{eq::k0t}\\
        2\frac{\overline{M}}{|\beta|}K_1(2\,|\overline{M}\beta|) 
        = {} & 0 + \mathcal{O}\left(m^2z_3^2,\frac{t^2 m^2}{z_E^2},
        1/\overline{z}^2\right) ,\label{eq::k1t}\\
        2\frac{\overline{M}^2}{\beta^2}K_2(2\,|\overline{M}\beta|) 
        = {} & 0 + 
        \mathcal{O}\left(m^2z_3^2,\frac{t^2 m^2}{z_E^2},1/\overline{z}^2\right)
        \, . \label{eq::k2t}
\end{align} 

Care must be taken when matching these expressions to the light-cone case. The
limits need to be taken in the right order so that the quantity $\frac{t^2
m^2}{z_E^2}$ remains small in the process. One must first consider the small
flow time regime at fixed $z_3$, in which $\overline{z}\gg 1$, and then consider
the limit in which $m^2 z_3^2$ goes to zero. Taking the limit of small $m^2
z_3^2$ at fixed $t$ would violate the condition above and invalidate the
factorization theorem, {\em viz.} data for values of $t$ and $z_3$ that
correspond to large values of $t^2 m^2/z_E^2$ are not described by the
factorization theorems discussed here. With this in mind, the only logarithmic
infrared divergence occurs in the first Bessel function, which has been expanded
using Eq.~\eqref{eq::besselexpansion}. Thus, in the small flow-time regime
Eq.~\eqref{eq:KtBessel} becomes
\begin{align}
        \label{eq:Ktlogz}
        K_t\left(-\overline{z}_3^2, \overline{M}^2\right) = {} & 
        \frac{1}{(4\pi)^3} 
        \left[-\log\left( M^2 z_3^2\right)+ 2 \log\left(2e^{-\gamma_E}\right) + 
                {\cal R}(Mz_3)\right] \nonumber \\
                & {} + \mathcal{O}\left(m^2z_3^2,
                        \frac{t^2 m^2}{z_E^2},1/\overline{z}^2\right)\,,
\end{align}
where the rational function ${\cal R}(Mz_3)$ contains the IR finite contributions 
generated by the $J$ functions of Eq.~\eqref{eq:J}.
The logarithmic IR divergence in Eq.~\eqref{eq:Ktlogz}, regularized by the mass $m$, matches those in Eqs.~\eqref{eq::Ioffe1loopren} and \eqref{eq::pIoffe1loopren}. 

In the short flow-time regime, the one-loop contributions to Eq.~\eqref{eq::MEflow} from diagrams (a) and (b) are just those given in Eq.~\eqref{eq::zmzphi}. The corresponding
renormalized quantity at one loop is therefore
\begin{align}
\label{eq::tIoffe1loopren}
    \widehat{\mathcal{M}}_t\left(\nu, -z_3^2;\,\mu^2\right) =&
    \left[1+ \alpha\left(\frac{1}{6}\log\frac{m^2}{\mu^2} + b\right)  \right]\widehat{\mathcal{M}}^{(0)}\left(\nu, 0\right) \nonumber \\
    &+\alpha \int_0^1 d\xi\,\left(1-\xi\right)
    \left[-\log\left( M^2 z_3^2\right)+ 2 \log\left(2e^{-\gamma_E}\right) + {\cal R}(Mz_3)\right] \widehat{\mathcal{M}}^{(0)}\left(x \nu, 0\right) .
\end{align}
We can now directly relate this quantity, via a factorization relation, to the light-cone quantity $f(x,\mu^2)$ using Eq.~\eqref{eq::PDFs}. We obtain
\begin{align}
	\label{eq::factt}
	\widehat{\mathcal{M}}_t\left(\nu, -z_3^2; \mu^2\right) = 
	\int_{0}^{1} dx\,\overline{C}\left(x\nu, \mu^2 z_3^2 \right) \widehat{f}\left(x,\mu^2\right)\, ,
\end{align}
with
\begin{align}
	\label{eq::Cpseudot}
	\overline{C}\left(x\nu,\mu^2 z_3^2 \right) = {} & 
	 e^{i x\nu} - \alpha\int_0^1 d\xi \, 
	 \left(1-\xi\right) \left[
	 \log\left( \mu^2 z_3^2\frac{e^{2\gamma_E}}{4} \right) - {\cal R}(Mz_3)\right] e^{i \xi x\nu} \nonumber \\
         & {} + \mathcal{O}\left(m^2 z_3^2,\frac{t^2 m^2}{z_E^2},
                1/\overline{z}^2\right)\, .
\end{align}
This factorization relation provides the explicit connection between the collinear PDFs and an
equal-time correlator at nonzero flow time, through a convolution with a perturbative kernel.

\section{Conclusions}
\label{sec:conclusion}

We have addressed the definition and renormalization of equal-time
correlators whose computation can be performed on the lattice, studying their
relation with the corresponding light-cone matrix elements underlying the
definition of collinear PDFs via factorization theorems.
To highlight and clarify the most important aspects of the factorization
theorems, we have studied them in the context of a nongauge theory. This allows
us to avoid the formal complications that arise in QCD, which can obscure the key concepts. 
We derive the relation between the light-cone and Euclidean matrix
elements at the one-loop level, and then study the limits that lead to well-defined
factorization theorems. These relations express suitable correlators that are evaluated by
Monte Carlo calculations in terms of a convolution between a collinear PDF and an
infrared safe coefficient function, which can be evaluated in perturbation
theory. We obtain factorization theorems in both position and momentum
space, by considering the regimes of small-$z_3^2$ and large-$P_3$ respectively, and
show that these limits are equivalent at one loop, which highlights the formal
equivalence of the pseudo- and quasi-PDFs approach. In addition, we demonstrate that
 the gradient flow can be used to define a new class of lattice observables that satisfy factorization.

These ideas naturally suggest that the lattice data should be used in a fitting framework to extract
PDFs, in the same way experimental data are usually included in global QCD
analyses. This approach has been studied at NLO in
\cite{Karpie:2019eiq,Cichy:2019ebf} and is in the spirit of the ``good lattice
cross-sections'' (or factorizable matrix elements) proposed in
Ref.~\cite{Ma:2017pxb,Ma:2014jla}. Results for the NNLO coefficients entering the factorization theorem have recently become available
in both position and momentum space \cite{Li:2020xml, Chen:2020ody, Braun:2020ymy, Chen:2020arf, Chen:2020iqi}, paving the way to NNLO fits.
 
The general idea is straightforward: the unknown $x$-dependence of the PDF at a
specific fitting scale is parametrized by introducing a suitable functional
form. The PDF at a generic scale can be computed in terms of its parametric form
at the fitting scale, which then leads to a theoretical prediction for the
lattice observable when working in either the small-$z_3^2$ or large-$P_3$
limit. Assuming that we have a set of lattice results for the real and imaginary
part of the Ioffe-time matrix elements, a standard minimum-$\chi^2$ fit yields
the values of the free parameters that best describe such data. As in any other
PDF determination, we highlight the importance of having a robust estimate of
the full covariance matrix that enters the $\chi^2$ definition, and this should
be provided by the lattice group performing the calculation.
 
We also stress that this procedure is exactly the one that is currently used to
extract PDFs from experimental data \cite{Ball:2017nwa, Dulat:2015mca,
Alekhin:2017kpj, Martin:2009iq, Buckley:2014ana}, with the lattice matrix
elements playing the same role as the cross-sections for high-energy processes.
Given a discrete set of points for quantities that are connected to collinear
PDFs through a factorization theorem, we can use them to perform a fit, thereby
obtaining an estimate of the PDFs and their corresponding error.
  
In this work we demonstrate, at one loop in a scalar model, the conceptual
equivalence of the pseudo and quasi distribution methods, and advocated for a
fitting framework that directly relates Ioffe time distributions to light-cone
PDFs. We emphasize, however, that conceptual equivalence may not translate to
equivalence in practice. On the one hand, the LaMET approach relies on large
hadronic momenta to suppress higher twist contamination. On the other hand, the
pseudo distribution approach uses small spatial separations to suppress higher
twist effects, but requires large momenta to cover a range of Ioffe times. In
both cases, large values of the hadron momentum can lead to significant
signal-to-noise challenges and discretization effects of the form $(aP)^n$. The
interplay of higher twist contamination and discretization effects is nontrivial
and will depend both on the details of the distribution itself and on the
specific choice of discretization. These effects must be studied systematically,
across a wide range of observables, to pin down systematic uncertainties and
strengthen the role that lattice QCD can play in the determination of hadron
structure.

\subsection*{Acknowledgments}
We are thankful to K.~Orginos, E.R.~Nocera, A.V.~Radyushkin, G.C.~Rossi, and M.~Testa for useful discussions.

TG is supported by The Scottish Funding Council, grant H14027. C.J.M. is
supported in part by the U.S. Department of Energy, Office of Science, Office of
Nuclear Physics under contract No. DE-AC05-06OR23177. LDD is supported by an
STFC Consolidated Grant, ST/P0000630/1, and a Royal Society Wolfson Research
Merit Award, WM140078.

\appendix

\section{Momentum space factorization}
\label{app:plus}

In this appendix we report in detail some of the computations performed in
Sec.~\ref{sec::momentumspace}, to obtain the coefficient $C$ of
Eq.~\eqref{eq::C0} and its high momentum limit of Eq.~\eqref{eq::matching}. In
order to compute the Fourier transform of the coefficient $\tilde{C}$ entering
Eq.~\eqref{eq::fact0}, we perform a change variable, $\theta = \xi P_3 z_3 $,
and define $\eta = \frac{y}{\xi}$, so that
\begin{align}
    \frac{P_3}{2\pi} 
    & \int_{-\infty}^{\infty} dz_3\,
        e^{-i y P_3 z_3}\,
        \tilde{C}\left(x P_3 z_3, m z_3, \frac{\mu^2}{m^2} \right)
        = \frac{1}{x} \int_{-\infty}^{\infty} \frac{d\theta}{2\pi}\,
        e^{-i \eta \theta}\,
        \tilde{C}\left(\theta, \frac{m \theta}{x P_3}, \frac{\mu^2}{m^2} \right)
        = \nonumber\\
    \label{eq:CFourierTransform}
    &= \frac{1}{x} \biggl[
        \delta\left(\eta-1\right) - \alpha \int_0^1 d\xi\, \left(1-\xi\right)\,
        \int_{-\infty}^{\infty} \frac{d\theta}{2\pi}\, 
        e^{-i \left(\eta-\xi\right) \theta} \,
        \biggl(2 K_0\left(\frac{M\theta}{x P_3}\right)
        -\log\frac{\mu^2}{M^2}\biggr)\biggr]\, .
\end{align}
The Fourier transform of the Bessel function, obtained also in Ref.~\cite{Radyushkin:2016hsy}, can be computed using the integral
representation in Eq.~\eqref{eq::Bessel}, computing the gaussian integral over
$\theta$ first:
\begin{align}
    \int_{-\infty}^{\infty} \frac{d\theta}{2\pi}\,
    & e^{-i \left(\eta-\xi\right)\theta}\,
    \int_0^{\infty} \frac{dT}{T}\, e^{-T} 
    e^{-\left(\frac{M\theta}{x P_3}\right)^2 \frac{1}{4T}} 
    =  \frac{1}{\sqrt{\left(\eta - \xi\right)^2 + \frac{M^2}{x^2 P_3^2}}}\, ,
\end{align}
so that the $\mathcal{O}(\alpha)$ contribution to \eqref{eq:CFourierTransform}
can be written as 
\begin{align}
    \label{eq::appC0}
    \int_0^1 d\xi\,
    \left(1-\xi\right) 
    \left[ \frac{1}{\sqrt{\left(\eta-\xi\right)^2 + \frac{M^2}{x^2P_3^2}}} 
    - \delta\left(\xi-\eta\right) \log\frac{\mu^2}{M^2}\right]\, .
\end{align}
As mentioned in Sec.~\ref{sec::momentumspace}, the computation of the
large-$P_3$ limit when $\eta \in \left(0,1\right)$ requires additional care,
since the integrand develops a non-integrable divergence for $\xi=\eta$ when
$M^2/(x^2P_3^2) \to 0$. This issue was first addressed and solved in Ref.~\cite{Radyushkin:2017lvu} in the context of QCD.
Since in the scalar theory the same kind of Bessel function appears, its Fourier transform leads to an analogous singularity. 
In order to elucidate this problem, given a generic test
function $\phi\left(\xi\right)$, we consider the integral
\begin{align}
    \label{eq::limitApp}
        \int_0^1 d\xi\,
        \frac{\phi\left(\xi\right)}{\sqrt{\left(\eta-\xi\right)^2 + \kappa^2}}
\end{align}
in the limit where $\kappa\to 0$. Defining 
\begin{equation}
    \label{eq:GfunDef}
    G\left(\eta, \kappa^2\right) = 
    \int_0^1 \frac{d\xi}{\sqrt{\left(\xi - \eta\right)^2 + \kappa^2}}
\end{equation}
allows us to rewrite Eq.~\ref{eq::limitApp} above as 
\begin{align}
    \label{eq:IntPlusDecomp}
    \int_0^1 d\xi\,
    \frac{\phi\left(\xi\right)}{\sqrt{\left(\eta-\xi\right)^2 + \kappa^2}}
    = \phi(\eta) G\left(\eta, \kappa^2\right) + 
    \int_0^1 d\xi\,
    \frac{1}{\sqrt{\left(\eta-\xi\right)^2 + \kappa^2}}
    \left(\phi\left(\xi\right)-\phi\left(\eta\right)\right)\, .
\end{align}
The divergence of the original integral is encoded in the function
$G\left(\eta,\kappa^2\right)$, which can be readily evaluated:
\begin{equation}
    \label{eq:GfunIntegral}
    G\left(\eta, \kappa^2\right) =
    \log \left(4\eta\left(1-\eta\right)\frac{1}{\kappa^2}\right) +
    \mathcal O\left(\kappa^2\right)\, .
\end{equation}
The integral on the RHS of \eqref{eq:IntPlusDecomp} is convergent for $\kappa\to
0$, and we have
\begin{align}
    \int_0^1 & d\xi\, 
    \frac{1}{\sqrt{\left(\eta-\xi\right)^2 + \kappa^2}} 
    \left(\phi\left(\xi\right)-\phi\left(\eta\right)\right) = 
    \nonumber \\
    &= \int_0^1 d\xi\, 
    \frac{1}{\left| \xi - \eta\right|}
    \left(\phi\left(\xi\right)-\phi\left(\eta\right)\right) 
    + \mathcal{O}(\kappa^2) \nonumber \\
    \label{eq:PlusDistrLimit}
    &= \int_0^1 d\xi\,
    \frac{1}{\left| \xi - \eta\right|_+}\,
    \phi\left(\xi\right) 
    + \mathcal{O}(\kappa^2) \, .
\end{align}
Therefore, collecting both contributions, 
\begin{align}
    \frac{1}{\sqrt{(\eta - \xi)^2 + \kappa^2}} = 
    \delta(\eta - \xi) \log\left(4\eta (1-\eta) \frac{1}{\kappa^2}\right)
    + \frac{1}{\left|\eta - \xi\right|_+} + \mathcal{O}\left(\kappa^2\right)\, .
\end{align}

\section{Equivalence between pseudo- and quasi-PDF approaches}
\label{app:check}

As discussed at the end of Sec.~\ref{sec:factorization}, taking the
small-$z_3^2$ limit in position space is equivalent to taking the large-$P_3$
limit in momentum space. This can be verified at 1-loop by showing that the
coefficent functions of Eqs.~\eqref{eq::Cpseudo0} and~\eqref{eq::matching} are
related through a Fourier transform, as stated in Eq.~\eqref{eq::check}. Here we
report the details of the computation. Taking the Fourier transform of the
small-$z_3^2$ coefficient of Eq.\eqref{eq::Cpseudo0} and defining $\eta= y/x$ we
have 
\begin{align}
    \label{eq::checkapp}
	\frac{P_3}{2\pi}&\int_{-\infty}^{\infty} dz_3\, e^{-i y P_3 z_3 }\,
	\tilde{C}\left(x\nu, \mu^2 z_3^2 \right) = 
    \frac{1}{x}\int_{-\infty}^{\infty}\frac{d\theta}{2\pi}\, e^{-i\theta\eta}\,
	\tilde{C}\left(\theta, \frac{\mu^2\theta^2}{x^2 P_3^2} \right) \nonumber \\
    & = \frac{1}{x}\biggl[\delta\left(\eta-1\right) 
    +\alpha\log\frac{4 \left(xP_3\right)^2}{\mu^2 e^{2\gamma_E}}\int_0^1 d\xi\,
    \delta\left(\xi-\eta\right)\left(1-\xi\right) \nonumber \\
    &\,\,\,\,\,\,\,\,\,\,\,\,\,\,\,\,\,\,\,\,\,\,\,\,\,\,\,\,\,\,
    -\alpha\int_0^1 d\xi \left(1-\xi\right)\int_{-\infty}^{\infty} \frac{d\theta}{2\pi} 
    e^{-i \left(\eta-\xi\right) \theta} \,\log\theta^2\biggr]\, .
\end{align}
Following Ref.~\cite{Izubuchi:2018srq}, the Fourier transform of $\log\theta^2$
can be defined as
\begin{align}
    \label{eq::FTlog}
    \int& \frac{d\theta}{2\pi} e^{-i t \theta} \log \theta^2
    = \left[\frac{d}{d\tau} \int \frac{d\theta}{2\pi} e^{-it \theta} 
    \left(\theta^2\right)^{\tau}\right]_{\tau=0} \nonumber\\
    &= -2 \gamma_E \, \delta\left(t\right) - \frac{\theta\left(1 - |t|\right)}{|t|_{\left(+0\right)}}
    - \frac{\theta\left(|t| -1\right)}{|t|_{\left(+\infty\right)}}
    + \frac{1}{\left(t\right)^2}\,\delta\left(\frac{1}{|t|}\right)\, ,
\end{align}
with
\begin{align}
    \label{eq::distribution1}
    &\frac{1}{|t|}_{\left(+0\right)} 
    = \lim_{a\rightarrow 0}
    \left[\frac{\theta\left(|t|-a\right)}{|t|} +  \delta\left(|t|-a\right)\log a\right], \\
    &\frac{1}{|t|}_{\left(+\infty\right)} 
    = \frac{1}{\left(t\right)^2}\lim_{a\rightarrow 0}
    \left[\theta\left(\frac{1}{|t|}-a\right)|t| + \delta\left(\frac{1}{|t|}-a\right)\log a\right], \\
    &\delta\left(\frac{1}{|t|}\right) = \lim_{a\rightarrow 0}\delta\left(\frac{1}{|t|} - a\right)\, .
\end{align}
The proof of Eq.~\eqref{eq::FTlog} can be found, for example, in the Appendix A
and C of Ref.~\cite{Izubuchi:2018srq}, to which we refer for more details.
Setting $t = \eta-\xi $ and plugging everything in Eq.~\eqref{eq::checkapp},
remembering that $\xi \in \left[0,1\right]$, we get different answers depending
on the value of $\eta$. For $\eta \in \left[0,1\right]$, just the first two
terms in Eq.~\eqref{eq::FTlog} contribute, giving
\begin{align}
    \label{eq::cont1}
    \int_0^1 &d\xi \left[2 \gamma_E \, \delta\left(\eta-\xi\right) -
    \lim_{a\rightarrow 0}
    \left(\frac{\theta\left(|\eta-\xi|-\beta\right)}{|\eta-\xi|} +  
    \delta\left(|\eta-\xi|-a\right)\log a\right)\right]
    \left(1-\xi\right) \nonumber \\
    & = \log{e^{2\gamma_E}}\left(1-\eta\right) + 
    \left(1-\eta\right)\log{\eta\left(1-\eta\right)} +2\eta -1\, ,
\end{align}
while for $\eta > 1$ or $\eta < 0$ the third contribution in Eq.~\eqref{eq::FTlog} gives simply
\begin{align}
    \label{eq::cont2}
    -\int_0^1 d\xi\, \left(1-\xi\right)\frac{|\eta-\xi|}{\left(\eta-\xi\right)^2}\, .
\end{align}
Looking at the last term in Eq.~\eqref{eq::FTlog}, considering its contribution
to the convolution integral with the PDF and doing the integral over $x$ first
we find
\begin{align}
    \label{eq::cont3}
    \lim_{a\rightarrow 0}\int_0^1 \frac{dx}{x}\int_0^1 d\xi \,\left(1-\xi\right)
    \delta\left(\frac{1}{|\frac{y}{x}-\xi|} - a\right)f\left(x\right) 
    \propto \lim_{a\rightarrow 0} a^2 f\left(a\right) = 0.
\end{align}
Using Eqs.~\eqref{eq::cont1}, \eqref{eq::cont2}, \eqref{eq::cont3} in
Eq.~\eqref{eq::checkapp} we find back the expression for
$C\left(\eta,\frac{\mu^2}{x^2 P_3^2}\right)$ as in Eq.~\eqref{eq::matching},
which completes our check.

\section{quasi-PDFs and their moments}
\label{app:moments}

As mentioned in the introduction of this paper, the works where the concept of
quasi-PDF was first introduced have been criticized in
Refs.~\cite{Rossi:2017muf, Rossi:2018zkn}, where it was argued that such
approach does not give access to the full nonperturbative PDF. In support of
their argument, the Authors have shown that moments of quasi-PDFs are divergent:
since the moments of parton distributions should reproduce the (finite) matrix
elements of the renormalized local DIS operator, they conclude that the
quasi-PDF cannot be considered as an euclidean generalization of the light-cone
PDF. The problem has been addressed in several independent papers, see e.g.
Refs.~\cite{Ji:2017rah, Radyushkin:2018nbf, Karpie:2018zaz}. In this appendix we
revise these criticisms in the framework of the scalar model: first we show how
the points raised in Ref.~\cite{Rossi:2017muf, Rossi:2018zkn} can be easily seen
and understood within the toy model presented in this paper, showing explicitly
how all the moments of quasi-PDFs are indeed divergent; second we discuss how
such feature does not invalidate the programme presented in
Sec.~\ref{sec:conclusion}, based on the determination of a parametric form of
the light-cone PDF based on a discrete set of data for the euclidean matrix
element.

We start this section by computing the moments of the quasi-PDF. From
Eq.~\eqref{eq::1loopcont}, using the integral representation of the Bessel
function, the $\mathcal{O}\left(\alpha\right)$ contribution to the euclidean
matrix element reads
 \begin{align}
    \hat{\mathcal{M}}^{(1)}\left(\nu, -z_3^2\right) = 
    \alpha\int_0^{1} d\xi \, \left(1-\xi\right) \int_0^{\infty}\frac{dT}{T} e^{-T} e^{-\frac{z_3^2 M^2}{4T}} e^{-i\xi P_3 z_3}\, .
\end{align}
The corresponding contribution to the quasi-PDF is found by taking the Fourier
transform of the expression above:
\begin{align}
    \hat{q}^{(1)}\left(y\right) &= 
    \frac{P_3}{2\pi}\int_{-\infty}^{\infty}d z_3\, e^{-i y P_3 z_3 } \hat{\mathcal{M}}^{(1)}\left(\nu, -z_3^2\right) \\
    & = \alpha\,\frac{P_3}{\sqrt{\pi}}\int_0^1 d\xi \, \left(1-\xi\right)\frac{1}{M} 
	\int_0^{\infty} \frac{dT}{\sqrt{T}}e^{-T} e^{-T\left(y+\xi\right)^2 \frac{P_3^2}{M^2}}\, ,
\end{align}
where in the last line we have computed the gaussian integral over $z_3$.
Taking the $n$-th moment of $\hat{q}^{(1)}\left(y\right)$ yields
\begin{align}
    \int_{-\infty}^{\infty} dy\, y^n \hat{q}^{(1)}\left(y\right) =
	\alpha\,\frac{P_3}{\sqrt{\pi}}\int_0^1 d\xi \, \left(1-\xi\right)
	\frac{1}{M} \int_0^{\infty} \frac{dT}{\sqrt{T}}e^{-T}
	\int_{-\infty}^{\infty} dy\, 
	\left(y-\xi\right)^n e^{-T y^2 \frac{P_3^2}{M^2}}\, .
\end{align}
We can expand the polynomial term as
\begin{align}
	\left(y-\xi\right)^n = \sum_{k=0}^{n}\binom{k}{n}y^{n-k}\xi^k\
\end{align}
and evaluate each contribution in turn. The term with $k=n$, performing the
integral over $y$ first, yields
\begin{align}
	 \alpha\,\frac{P_3}{\sqrt{\pi}}&\int_0^1 d\xi\, 
	 \left(1-\xi\right) \xi^n \frac{1}{M}
	 \int_0^{\infty} \frac{dT}{\sqrt{T}}\, e^{-T } 
	 \int_{-\infty}^{\infty} dy\, e^{-T y^2 \frac{P_3^2}{M^2}}  \\
	 & = \alpha \int_0^1 d\xi\, 
	 \left(1-\xi\right) \xi^n \int_0^{\infty} \frac{dT}{T}e^{-T}\, . 
\end{align}
The integral over $T$ is divergent, with the divergence originating from the
lower end of the integration region, i.e. when $T\to 0$. Introducing a cutoff
$a^2$ for small values of $T$~\footnote{The cutoff $a$ has dimensions of length
and can be thought of as a lattice spacing if the theory were regulated on a
lattice.} and considering the limit $a^2 \rightarrow 0$, we get the logarithmic
divergent contribution
\begin{align}
\label{eq::logdiv}
	\alpha\int_0^1 d\xi\, 
	\left(1-\xi\right) \xi^n \int_{a^2}^{\infty} \frac{dT}{T}e^{-T}\,\,
	\stackrel{a^2\rightarrow 0}{\sim}\,\, 
	- \alpha\int_0^1 d\xi\, \left(1-\xi\right) \xi^n \log a^2\, .
\end{align}
Similarly we can consider contributions coming from even values of $n-k$. Using 
\begin{align}
	\int_{-\infty}^{\infty}
	&dy\, y^{2m} e^{-T y^2 \frac{P_3^2}{M^2}} = 
		\frac{M}{P_3} \left(-\frac{M^2}{P_3^2}\frac{d}{dT}\right)^m 
		\int_{-\infty}^{\infty} dy\,e^{-T y^2} \nonumber \\
	&= \frac{M\sqrt{\pi}}{P_3} 
	\left(-\frac{M^2}{P_3^2}\frac{d}{dT}\right)^m \frac{1}{\sqrt{T}}  
	\propto \frac{M \sqrt{\pi}}{P_3} \frac{1}{T^{m + \frac{1}{2}}}\, ,
\end{align}
and considering $n-k = 2m$, we get
\begin{align}
	\label{eq::powerdiv}
	\alpha\, \frac{P_3}{\sqrt{\pi}}
	&\int_0^1 d\xi\, \left(1-\xi\right) \xi^{n-2m} \frac{1}{M}
		\int_0^{\infty} \frac{dT}{\sqrt{T}}\, e^{-T } 
		\int_{-\infty}^{\infty} dy\, y^{2m} e^{-T y^2 \frac{P_3^2}{M^2}} \nonumber \\
	&\propto \alpha\, \int_0^1 d\xi\, \left(1-\xi\right) \xi^{n-2m}
		\int_{a^2}^{\infty} \frac{dT}{T^{m +1}}\, e^{-T} \,\, 
		\stackrel{a^2\rightarrow 0}{\sim}\,\, 
		\alpha\, \int_0^1 d\xi\, \left(1-\xi\right) \xi^{n-2m}
		\frac{1}{m} \left(\frac{1}{a^2} \right)^{m}\, ,
\end{align}
where again we have introduced a cutoff $a^2$ for small values of $T$ and
considered the limit $a^2 \rightarrow 0$. Contributions from odd values of $n-k$
vanish. Looking at Eqs.~\eqref{eq::logdiv}, \eqref{eq::powerdiv} it is then
clear that all the moments of the quasi-PDFs will be at least logarithmically
divergent with the cutoff $a^2$, with higher moments affected by higher power
divergences.

This relatively simple calculation shows that we obtain divergent contributions
for the moments of the quasi-PDF and therefore quasi-PDFs cannot be considered
as the proper euclidean generalization of the light-cone parton distribution.
This, however, does not invalidate the approach described in
Sec.~\ref{sec:conclusion}: as mentioned, what really matters is the existence
of a factorization theorem connecting the collinear PDF with a renormalizable
quantity that can be computed on the lattice, which in our case will be the
euclidean matrix element of Eq.~\eqref{eq::pIoffe1loopren}, computed for fixed
values of $P_3$ and $z_3$. As long as $z_3$ is kept small and different from
$0$, the factorization formula \eqref{eq::fact2} holds, and can be used to fit
the light-cone PDF using the available lattice data. How well such data can
constrain the PDF is something which should be investigated, just as in the same
way the constraints from new experimental measurements are usually analyzed.

\bibliographystyle{UTPstyle}
\bibliography{main}

\end{document}